\documentclass[aps,12pt,showpacs]{revtex4}
\usepackage{graphics}
\usepackage{psfig}
\def\beq{\begin{equation}}
\def\eeq{\end{equation}}

\def\bea{\arraycolsep .1em \begin{eqnarray}}
\def\eea{\end{eqnarray}}
\def\Tr{{\rm Tr}}
\def\step{\\[-1.5ex]}

\def\eq#1{(\ref{#1})}

\def\s0#1#2{\mbox{\small{$ \frac{#1}{#2} $}}}
\def\0#1#2{\frac{#1}{#2}}


\begin{document}

{\normalsize\begin{flushright}
CERN-TH-2003-247\\
USACH-2003-012\\[15ex] \end{flushright}
}
\title{Subleading critical exponents from the renormalisation group\\[5ex]
}

\author{Daniel F. Litim}
\email{Daniel.Litim@cern.ch}
 \affiliation{Theory Division, CERN, CH-1211 Geneva 23.
}%

\author{Lautaro Vergara} \email{lvergara@lauca.usach.cl}
\affiliation{\mbox{Departamento de F\'{\i}sica, Universidad de
Santi\-ago de Chile,}\\ Casilla 307, Santi\-ago 2, Chile.  }%
\author{{}}

\begin{abstract}
${}$\\[3ex]

\centerline{\bf Abstract} We study exact renormalisation group
equations for the $3d$ Ising universality class. At the Wilson-Fisher
fixed point, symmetric and anti\-symmetric correction-to-scaling
exponents are computed with high accuracy for an optimised cutoff to
leading order in the derivative expansion. Further results are
derived for other cutoffs including smooth, sharp and background
field cutoffs.  An estimate for higher order corrections is given as
well.  We establish that the leading anti\-symmetric corrections to
scaling are strongly subleading compared to the leading symmetric
ones.
\end{abstract}


\maketitle

\newpage
\pagestyle{plain} \setcounter{page}{1} \noindent

\pagestyle{plain}
\setcounter{page}{1}
\noindent
{\bf 1. Introduction}\step

Many physical systems with short range interactions and a scalar order
parameter display Ising universal behaviour close to the critical
point. Initially introduced for the study of magnetic systems, the
Ising model also describes the physics of the liquid-gas phase
transition, transitions in binary mixtures and in Coulombic systems
\cite{Pelissetto:2000ek}. In high energy physics, Ising universal
behaviour is expected in various theories including the QCD phase
transition with finite quark masses \cite{Pisarski:ms}, the chiral
phase transition of QCD \cite{Pisarski:ms,Gavin:yk}, and
the electro-weak phase transition \cite{Rummukainen:1998as}.\step
 
The original Ising model has a global $Z_2$ symmetry. However, many
systems in the Ising universality class like the liquid-gas and the
electro-weak phase transition do not possess the $Z_2$ symmetry away
from the critical point. The presence of operators odd under $Z_2$
lead to additional corrections-to-scaling exponents. In principle,
deviations from $Z_2$ symmetric scaling are detectable
experimentally. Anti\-symmetric corrections to scaling $\sim L^{-0.5}$
have been revealed in a Monte-Carlo simulation of the electro-weak
phase transition \cite{Rummukainen:1998as}.  Previous theoretical
studies of anti\-symmetric corrections to scaling are based on the
$\epsilon$-expansion \cite{wegner}, the scaling field method
\cite{ScalingField}, and the Wegner-Houghton equation
\cite{Tsypin:2001ix}.\step

In this Letter, we study corrections to scaling for the $3d$ Ising
universality class using the Exact Renormalisation Group, which is
based on the Wilsonian idea of successively integrating out momentum
modes (see \cite{Bagnuls:2000ae} for reviews and references therein).
This approach is implemented through an infra-red cutoff which, within
a few constraints, can be chosen freely.  The strengths of the method
are its flexibility and its numerical stability.  Furthermore, a
general optimisation procedure is available, enabling a choice of the
infra-red cutoff best suited for the problem at hand
\cite{Litim:2000ci}.  To leading order in a derivative expansion, we
employ an optimised cutoff and compute the first six subleading
corrections-to-scaling exponents with high accuracy. We also obtain
results for smooth, sharp and background field cutoffs, and estimate
higher order corrections.\\[3ex]

\noindent
{\bf 2. Renormalisation group and critical exponents}\step

Renormalisation group methods have been used very successfully in the
computation of universal observables at second order phase
transitions. A particularly useful approach is the Exact
Renormalisation Group (ERG), based on the Wilsonian idea of
integrating out momentum modes within a path integral representation
of quantum field theory \cite{Bagnuls:2000ae}.  In its modern form,
the ERG flow for an effective action $\Gamma_k$ for bosonic fields
$\varphi$ is given by the simple one-loop expression
\beq \label{ERG}
\partial_t\Gamma_k[\varphi] =
\frac{1}{2}\Tr\left( \Gamma^{(2)}_k+ R\right)^{-1}    \partial_t  R.
\eeq
Here, $t\equiv\ln k$ is the logarithmic scale parameter, the trace
denotes a momentum trace and a sum over indices,
$\Gamma^{(2)}[\varphi](p,q)\equiv
\delta^2\Gamma/\delta\varphi(p)\delta\varphi(q)$, and $R$ is an
infra-red momentum cutoff at the momentum scale $k$.  The flow
\eq{ERG} interpolates between an initial (microscopic) action in the
ultra-violet and the full quantum effective action in the infra-red.
At every momentum scale $k$, \eq{ERG} receives its main contributions
for momenta about $p^2\approx k^2$.  The regulator can be chosen
freely and allows for an optimisation of the flow within general
approximations \cite{Litim:2000ci}.  The optimisation entails improved
convergence and stability properties of the flow.  In combination with
the numerical stability of the flow, it provides the basis for
reliable predictions based on systematic approximations to \eq{ERG}.
An important non-perturbative approximation scheme is the derivative
expansion \cite{Golner:1986}. To leading order, the local potential
approximation consists in the Ansatz
\beq\label{AnsatzGamma} \Gamma_k=\int d^dx \left(U_k(\varphi)
             + \012 \partial_\mu \varphi\partial_\mu \varphi \right)
\eeq
for the effective action. It implies that higher order corrections
proportional to the anomalous dimension $\eta$ of the fields are
neglected. For the Ising universality class, $\eta$ is of the order of
a few percent. Inserting the Ansatz \eq{AnsatzGamma} into the flow
equation \eq{ERG} and evaluating it for constant fields leads to the
flow for the effective potential $U_k$. We introduce dimensionless
variables $u(\phi)=U_k/k^d$ and $\phi= \varphi k^{1-d/2}$. Then,
finding a fixed point amounts to solving $\partial_t u=0$. To that end
we employ a polynomial truncation of the fixed point potential, retaining
vertex functions $\phi^{n}$ up to a maximum number $n_{\rm trunc}$,
\beq\label{PolyAnsatz0} u(\phi)
=
\sum^{n_{\rm trunc}}_{n=1}
 \s01{n!} \      \tau_n\ \phi^{n}\,.
\eeq
The potential has been normalised as $u(\phi=0)=0$. The Ansatz
\eq{PolyAnsatz0} leads to $n_{\rm trunc}$ ordinary differential
equations $\partial_t\tau_i\equiv \beta_i$. In three Euclidean
dimensions, the flow equation exhibits the nontrivial Wilson-Fisher
fixed point $u_*\neq 0$. Universal critical exponents and
corrections-to-scaling exponents are obtained as the
eigenvalues of the stability matrix at criticality
$M_{ij}=\partial\beta_{i}/\partial\tau_j|_*$.  For convenience, we
introduce the set of $\phi$-even couplings $\lambda_n=\tau_{2n}$ and
$\phi$-odd couplings $\bar\lambda_n=\tau_{2n-1}$. Under reflection in
field space the couplings and their $\beta$-functions behave as
\beq\label{reflection}
\phi\to
-\phi:\quad\left\{
\begin{array}{ccc}
(\lambda,\bar \lambda)
&\to 
& (\lambda,-\bar \lambda)
\\
(\beta_\lambda,\beta_{\bar \lambda})
&\to 
&(\beta_\lambda,-\beta_{\bar \lambda})
\end{array}\right.
\eeq
The scaling solution is symmetric under $\phi\to -\phi$. Hence all
$\phi$-odd couplings $\bar\lambda_*$ vanish at the fixed point. The
computation of critical exponents is simplified by observing that
$\beta_{\bar \lambda}(\lambda,\bar\lambda)=-\beta_{\bar
\lambda}(\lambda,-\bar\lambda)$ for all $\lambda$. This follows from
\eq{PolyAnsatz0} and \eq{reflection}. In particular, $\beta_{\bar
\lambda}$ vanishes identically at $\bar \lambda=0$, where all
derivatives of $\beta_{\bar\lambda}$ w.r.t.~the symmetric couplings
vanish at the fixed point. Therefore, the stability matrix $M$ at
criticality simplifies and becomes equivalent to the matrix
\begin{equation}
\label{stabilitymatrix}
\left(
\begin{array}{cc}
A&B\\ 0 &C
\end{array}
\right)
\end{equation}
with $A\equiv\partial \beta_{\lambda}/\partial\lambda|_*$,
$B\equiv\partial \beta_{\lambda}/\partial\bar \lambda|_*$ and
$C\equiv\partial \beta_{\bar \lambda}/\partial\bar\lambda|_*$\,.  In
consequence, the eigenvalues of $M$ reduce to those of the sub-matrices
$A$ and $C$. The matrix $A$ carries the information about the critical
exponent $\nu$ and the symmetric corrections to scaling, while the
matrix $C$ contains the information about anti\-symmetric corrections
to scaling.\\[3ex]

\begin{center}
\begin{figure}
\unitlength0.001\hsize
\begin{picture}(400,550)
\put(200,50){\large $n_{\rm trunc}$}
\put(300,160){\Large $-\nu^{-1}$}
\put(300,270){\Large ${}\ \ \omega$}
\put(300,400){\Large ${}\ \ \omega_2$}
\psfig{file=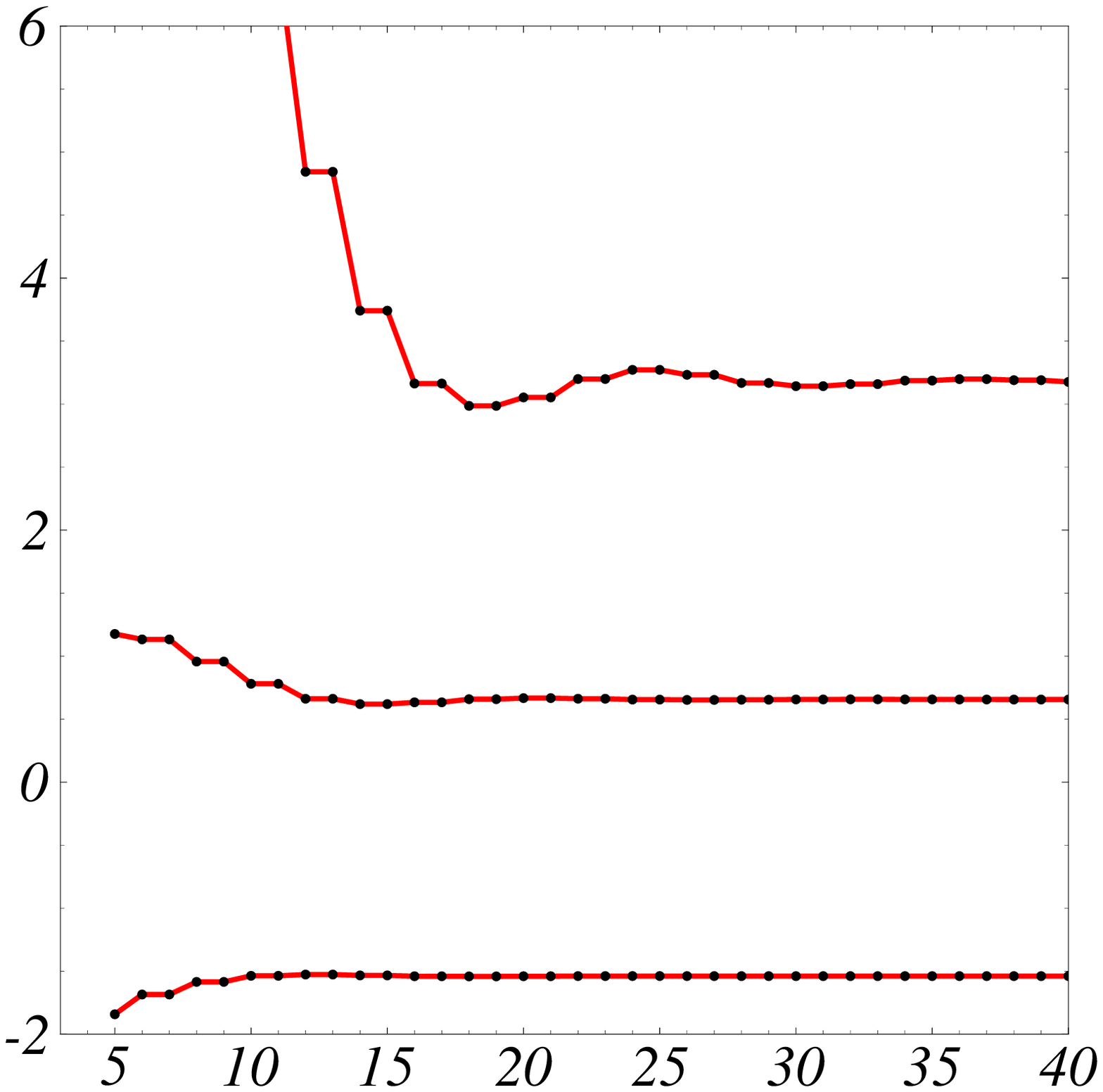,width=.45\hsize}
\end{picture}
\vskip-.1cm
\begin{minipage}{.58\hsize}
 \small {\bf Figure 1:} The exponents $\nu, \omega$ and $\omega_2$
 from a poly\-nomial truncation to order $n_{\rm trunc}$ about
 vanishing field.\\
\end{minipage}
\end{figure}
\end{center}

\noindent
{\bf 3. Results}\step

In this section, we present our results for the universal eigenvalues
of the stability matrix. The fixed point is determined in truncations
up to $n_{\rm trunc}=40$. The optimised regulator $R_{\rm
opt}=(k^2-q^2)\theta(k^2-q^2)$ is employed to improve the convergence
and stability of the flow
\cite{Litim:2000ci,Litim:2001dt,Litim:2002cf}. The stability matrix is
evaluated with two different methods: an expansion in powers of the
field about vanishing field, and an expansion in Legendre
polynomials. The latter case involves an integration in field space.
\step

Our numerical results are given in Tab.~1 and 2, and in Figs.~1-4.
First, we discuss our results for the $\phi$-even corrections to
scaling. In Fig.~1, the three leading eigenvalues of $A$ are
displayed, $\nu^{-1}$, $\omega$ and $\omega_2$. Notice that the
eigenvalues are identical for truncations $(n,n+1)$, if $n$ is
even. The reason for this is simple: increasing the truncation by a
$\phi$-odd coupling does neither change the dimension of the matrix
$A$, nor the numerical values of the fixed couplings (because
$\phi$-odd couplings vanish at the fixed point). Hence, the
$\phi$-even eigenvalues remain unchanged, as is clearly seen in
Fig.~1. The numerical values for the $\phi$-even eigenvalues are
identical to those which are found in a polynomial expansion in
$\rho=\phi^2/2$.

\begin{center}
\begin{figure}
\unitlength0.001\hsize
\begin{picture}(400,550)
\put(200,50){\large $n_{\rm trunc}$}
\put(300,450){\framebox{\huge ${}\  \bar \omega^{{}^{}}_1\ $}}
\psfig{file=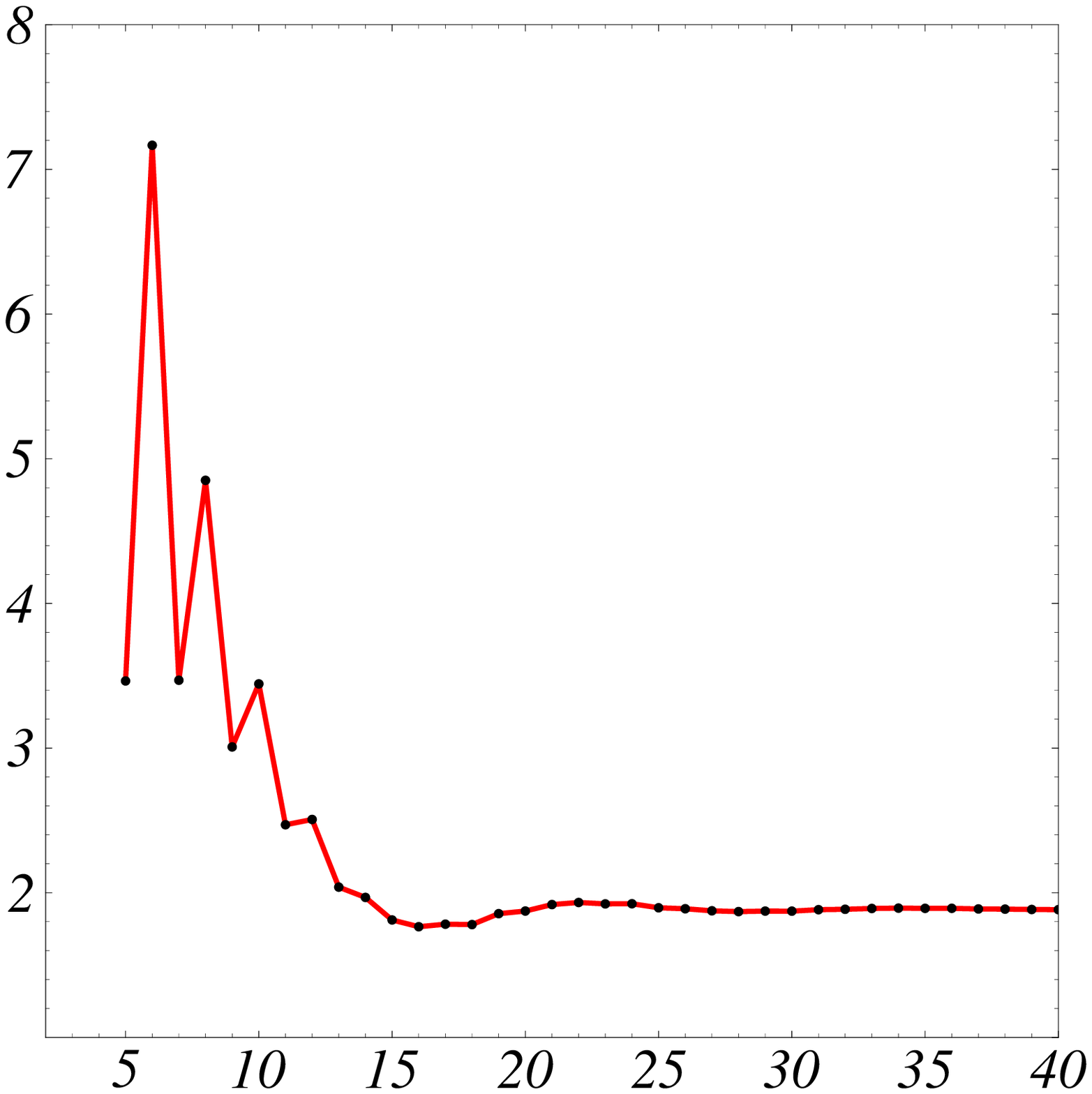,width=.45\hsize}
\end{picture}
\vskip-.5cm
\begin{minipage}{.6\hsize}
  { \small {\bf Figure 2:} The exponent $\bar \omega_1$ (see Fig.1).}
\end{minipage}
\end{figure}
\end{center}

\begin{center}
\begin{tabular}{cl|cl}
\multicolumn{2}{c|}{$\quad{} \phi$-even $\quad{}$ }
&\multicolumn{2}{c}{$\quad{} \phi$-odd $\quad{}$ }\\[1ex]\hline
&
&$\quad{}    y_h   \quad{}    $
&$\quad{}    -2.5   \quad{}    $
\\
$\quad{}     \nu   \quad{}    $ 
&$\quad{} 0.6495\quad{}    $
&$\quad{}     y_{\rm shift}  \quad{}    $ 
&$\quad{}     -0.5  \quad{}    $
\\
$\quad{}     \omega_1   \quad{}    $ 
&$\quad{}  0.655 \quad{} $
&$\quad{}     \bar\omega_1   \quad{}    $ 
&$\quad{} 1.88\quad{} $
\\
$\quad{}     \omega_2   \quad{}    $ 
&$\quad{}  3.18  \quad{} $
&$\quad{}     \bar\omega_2   \quad{}    $ 
&$\quad{} 4.5\quad{} $
\\
$\quad{}     \omega_3   \quad{}    $ 
&$\quad{}   5.9\quad{} $
&$\quad{}     \bar\omega_3   \quad{}    $ 
&$\quad{} 7\quad{} $
\end{tabular}
\end{center}
\begin{center}
\begin{minipage}{.65\hsize}
  \vskip.1cm {\small {\bf Table 1:} $\phi$-even and $\phi$-odd
    eigenvalues ($R_{\rm opt}$, $n_{\rm trunc} = 40$).\\[1ex]}
\end{minipage}
\end{center}

Next, we turn to the $\phi$-odd corrections to scaling (Figs.~2-4). We
have computed the eigenvalues of the matrix $C$ for $n_{\rm trunc}$ up
to $n_{\rm trunc}=40$. We find two eigenvalues $y_h=-5/2$ and $y_{\rm
shift}=-1/2$, related to redundant operators \cite{ScalingField}. All
other eigenvalues are positive. We denote them as $\bar\omega_n$. The
leading non-trivial $\phi$-odd correction-to-scaling exponent
$\bar\omega_1$ as a function of the truncation is displayed in
Fig.~2. (In the literature, $\bar\omega_1$ is sometimes denoted as
$\omega_A$ or $\omega_5$.)  Our results for $\bar\omega_2$ and
$\bar\omega_3$ are given in Fig.~3 and 4, respectively. Notice that
the pattern of the results, with increasing truncation, is similar to
the pattern found in the $\phi$-even sector. The results for two
subsequent truncations $(n,n+1)$ for $n$ odd are close to each other
for sufficiently large truncation. The reason for this is the
following: increasing the truncation by a $\phi$-even coupling does
not change the dimension of the matrix $C$. However, it does change
the numerical value of the fixed point, and hence the eigenvalues of
$C$. With increasing truncation, the numerical change within the
$\phi$-even couplings at the fixed point is very small and eventually,
the eigenvalues of $C$ become insensitive to the addition of a
$\phi$-even coupling. This is nicely observed in the results presented
in Figs.~2-4. Comparing the symmetric with the anti\-symmetric
corrections to scaling, the general pattern we find is that
$0<\omega_1<\bar\omega_1<\cdots< \omega_n<\bar\omega_n<\cdots$.

\begin{figure}[t]
\begin{center}
\unitlength0.001\hsize
\begin{picture}(1000,550)
\put(300,450){\framebox{\huge ${}\  \bar \omega^{{}^{}}_2\ $}}
\put(800,450){\framebox{\huge ${}\  \bar \omega^{{}^{}}_3\ $}}
\put(220,50){\large $n_{\rm trunc}$}
\put(740,50){\large $n_{\rm trunc}$}
\psfig{file=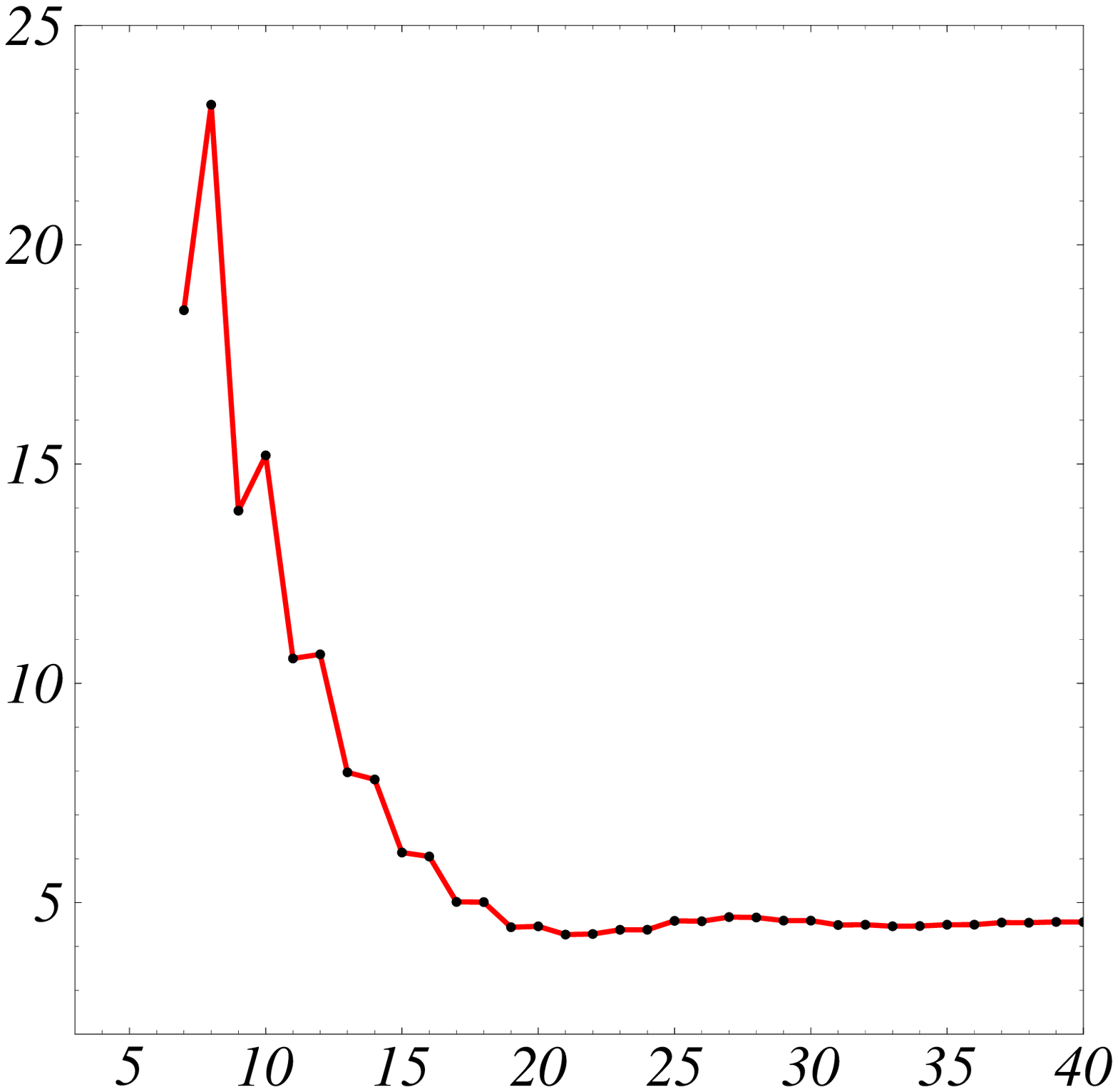,width=.45\hsize}
\hskip.05\hsize
\psfig{file=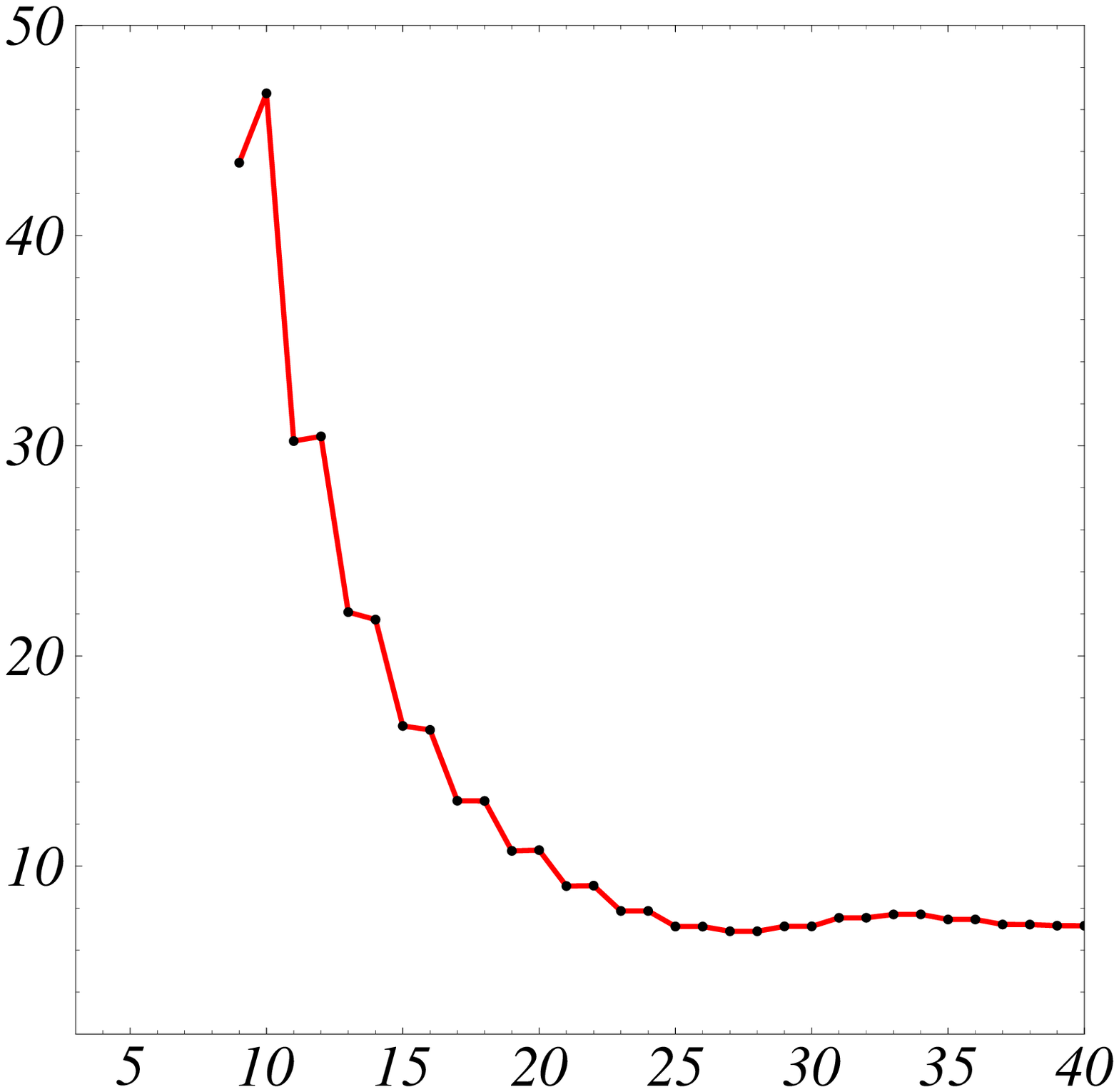,width=.45\hsize}
\end{picture}
\vskip-.5cm
\begin{minipage}{.47\hsize}
  { \small {\bf Figure 3:} The exponent $\bar \omega_2$ (see Fig.1).}
\end{minipage}
\hskip.05\hsize
\begin{minipage}{.47\hsize}
  { \small {\bf Figure 4:} The exponent $\bar \omega_3$ (see Fig.1).}
\end{minipage}
\end{center}
\end{figure}

\begin{center}
\begin{tabular}{cl|cl}
\multicolumn{2}{c|}{$\quad{} \phi$-even $\quad{}$ }
&\multicolumn{2}{c}{$\quad{} \phi$-odd $\quad{}$ }\\[1ex]\hline
& &$\quad{}    y_h   \quad{}    $ &$\quad{}    -2.5   \quad{}    $
\\
$\quad{}     \nu   \quad{}    $ &$\quad{} 0.649562\quad{}    $
&$\quad{}     y_{\rm shift}  \quad{}    $ &$\quad{}     -0.5  \quad{} $
\\
$\quad{}     \omega_1   \quad{}    $ &$\quad{}  0.6557 \quad{} $
&$\quad{}     \bar\omega_1   \quad{}    $ &$\quad{} 1.886\quad{} $
\\
$\quad{}     \omega_2   \quad{}    $ &$\quad{}  3.180  \quad{} $
&$\quad{}     \bar\omega_2   \quad{}    $ &$\quad{} 4.524\quad{} $
\\
$\quad{}     \omega_3   \quad{}    $ &$\quad{}   5.912\quad{} $
&$\quad{}     \bar\omega_3   \quad{}    $ &$\quad{} 7.33\quad{} $
\end{tabular}
\end{center}
\begin{center}
\begin{minipage}{.75\hsize}
  \vskip.1cm {\small {\bf Table 2:} $\phi$-even and $\phi$-odd
    eigenvalues ($R_{\rm opt}$, $n_{\rm trunc} = 22$, $\phi_{\rm
    max}=0.46$).\\[1ex]}
\end{minipage}
\end{center}

We have also computed the critical exponents by using the approach of
\cite{Tsypin:2001ix}. Expanding the scaling potential and the
eigenperturbations in terms of orthogonal polynomials (Legendre
polynomials) implies that the matrix elements of \eq{stabilitymatrix}
involve an integration in field space $\phi\in [-\phi_{\rm
max},\phi_{\rm max}]$. The weak dependence on $\phi_{\rm max}$ is
fixed by requiring that the $\phi$-even eigenvalues agree to high
accuracy with the known values obtained in \cite{Litim:2002cf} using
an expansion in $\rho=\phi^2/2$ about the potential minimum
$\rho=\rho_0$. This procedure improves the numerical convergence. Our
results for the eigenvalues are given in Tab.~2. They are consistent
with and have a higher accuracy than those given in Tab.~1.  \step

\begin{center}
\begin{figure}
\unitlength0.001\hsize
\begin{picture}(400,550)
\put(200,50){\large $n_{\rm trunc}$}
\put(300,450){\framebox{\huge ${}\  \bar \omega^{{}^{}}_1\ $}}
\put(151,202){\large ${}_{\rm quartic}$}
\put(60,470){\large ${}_{\rm sharp}$}
\put(120,255){\large ${}_{\rm opt}$}
\put(90,320){\large ${}_{\rm bg}$}
\psfig{file=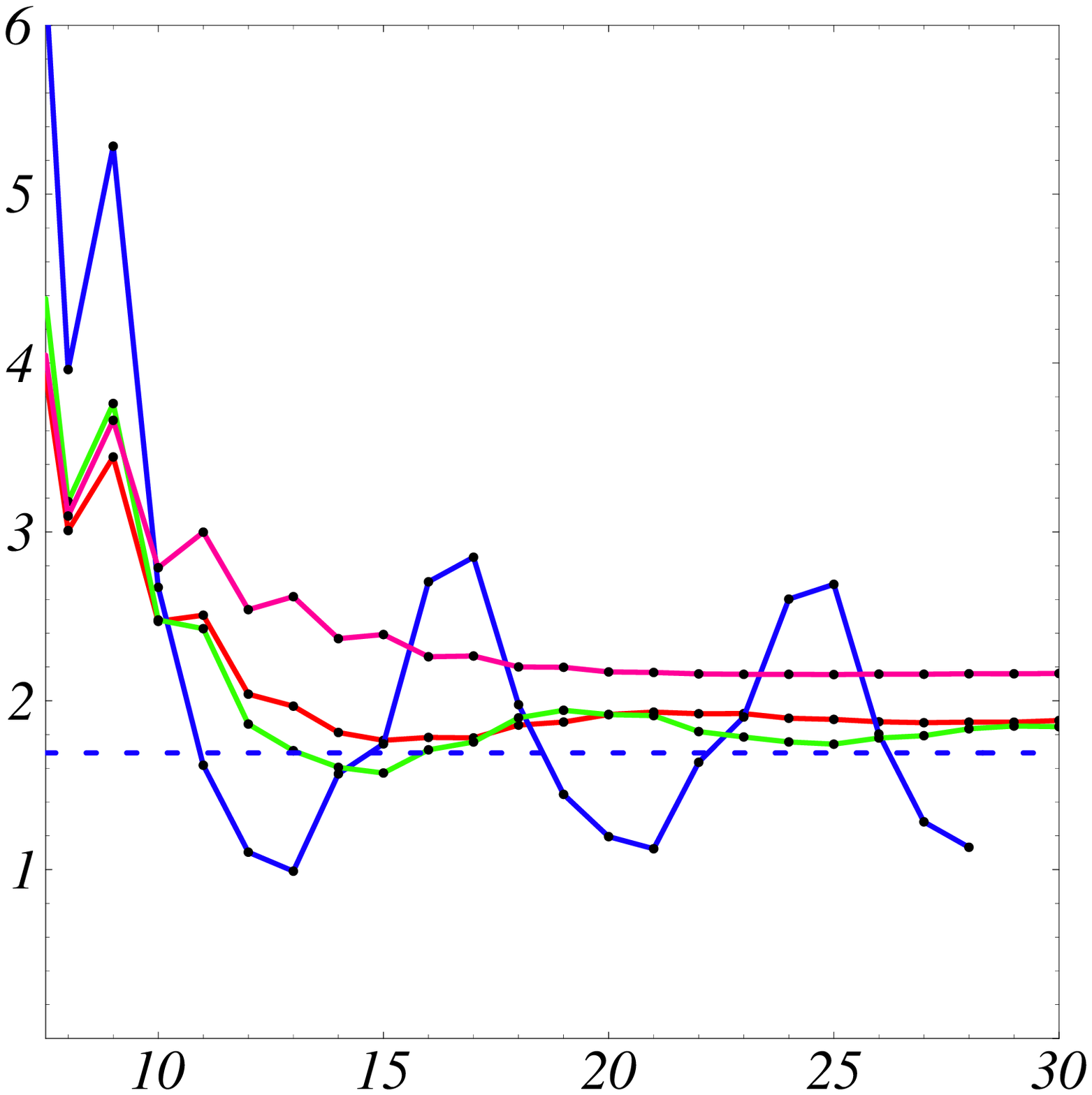,width=.45\hsize}
\end{picture}
\end{figure}
\vskip-.5cm
\begin{minipage}{.65\hsize}
{ \vskip-.5cm \small {\bf Figure 5:} The exponent $\bar\omega_1$ from a
  polynomial expansion up to order $n_{\rm trunc}$, and in comparison
  for the sharp, the back\-ground field (bg), the optimised (opt) and
  the quartic cutoff (see text).\\ {}}
\end{minipage} 
\end{center}

Next we discuss our results based on other regularisations including
the power law cutoff, the sharp cutoff and a background field
cutoff. Varying the momentum dependence of the regulator $R$ from
``smooth'' to ``sharp'' allows for an estimate of higher order
corrections due to operators neglected in the present approximation,
$e.g.$~\cite{Freire:2000sx}. We have computed the $\phi$-even and
$\phi$-odd eigenvalues for a smooth power-like regulator $R_{\rm
quartic}=k^2\cdot (q^4/k^4)$, for the sharp cutoff $R_{\rm
sharp}=\lim_{a\to\infty}a\,k^2 \theta(k^2-q^2)$, and for a background
field cutoff $R_{\rm bg}$. Results are summarised in Fig.~5 and
Tab.~3.\step

The power-law cutoff $R_{\rm quartic}$ is optimised in the sense of
\cite{Litim:2000ci}. We find that the flow based on $R_{\rm quartic}$
has similar stability and convergence properties as the flow with
$R_{\rm opt}$, $e.g.$~Fig.~5. Also, the numerical results as given in
Tab.~3 are within 5\% or less to each other.  The sharp cutoff does
not lead to an optimised flow \cite{Litim:2000ci}. It displays
instabilities within a local polynomial expansion about vanishing
field \cite{Margaritis:1987hv,Litim:2002cf}. This is confirmed in our
analysis. In Fig.~5, the $\phi$-odd eigenvalue $\bar\omega_1$ is
displayed up to a truncation $n_{\rm trunc}=30$.  The local field
expansion based on $R_{\rm sharp}$ oscillates in the eight-fold
pattern $(++++----)$ about its mean value, reminiscent of the
four-fold pattern in an expansion in $\phi^2$,
\cite{Litim:2002cf}. The expansion fails to converge at the present
order. The asymptotic value for $\bar \omega_1$ is indicated by the
dashed line. Moreover, the eigenvalues for $n_{\rm trunc}=16, 17, 24$
and 25 possess a small imaginary part, which is not displayed in
Fig.~5. These findings are a reflection of an intrinsic instability of
the sharp cutoff flow. Furthermore, the sharp cutoff value for the
leading critical exponent $\nu$ is $\sim 8\%$ larger than the value
for $R_{\rm opt}$ \cite{Litim:2001dt}, and $\sim 10\%$ larger than the
physical value. These properties indicate that quanti\-tative results
from the sharp cutoff flow within a given truncation, although
qualitatively correct, are less reliable than those by optimised
cutoffs.\step

The instability in the stability matrix of the sharp cutoff flow is
removed by expanding the fixed point potential and the
eigenperturbations in terms of Legendre polynomials. For the
eigenvalues, we adjust $\phi_{\rm max}$ as described above to improve
the numerical convergence.  Our results are given in Tab.~3, and by
the dashed line in Fig.~5. In the $\phi$-even sector, our results
agree to all significant digits with those by Morris (quartic cutoff)
\cite{Morris:1994ie}, Comellas and Travesset (sharp cutoff)
\cite{Comellas:1997tf} and Litim (optimised cutoff)
\cite{Litim:2002cf}. This provides a non-trivial consistency check,
because the numerical methods employed in
\cite{Morris:1994ie,Comellas:1997tf}, in \cite{Litim:2002cf}, and
here, are all different.  The main new results concern the eigenvalues
in the $\phi$-odd sector, where we also confirm the first two
eigenvalues by Tsypin (sharp cutoff) \cite{Tsypin:2001ix} (see also
\cite{BBS}). \\

\begin{center}
\begin{tabular}{r|l|l|l|l||r|l|l|l|l}
$\phi$-even${}\ $ 
&$\ {}R_{\rm sharp}$
&$\, {}R_{\rm quartic}$
&$\ \ \ {}R_{\rm opt}$
&$\ \ \ {}R_{\rm bg}$
&$\ \phi$-odd${}\ $ 
&$\ {}R_{\rm sharp}$
&$\, {}R_{\rm quartic}$
&$\ \ {}R_{\rm opt}$
&$\ \ {}R_{\rm bg}$
\\[.5ex]\hline
 ${}     \nu\ \ \        $ 
&$\ {}  0.6895\  $
&$\ {}  0.6604\  {} $ 
&$\ {}  0.649562 \       {}    $ 
&$\ {}  0.625979\       {}    $ 
&
&
&
&
\\
${}     \omega_1\ \ \       $ 
&$\ {}  0.5952\ $
&$\ {}  0.6285\ {} $
&$\ {}  0.6557\ {}    $ 
&$\ {}  0.762204\ {}    $ 
&$\quad{} \bar\omega_1\ \ \  $
&$\ {}   1.691\ $
&$\ {}   1.812\ {}    $ 
&$\ {}   1.886\ {}$ 
&$\ {}   2.163\ {}$ 
\\
$\quad{}     \omega_2\ \ \        $ 
&$\ {}   2.838\ $ 
&$\ {}   3.048\  {}    $
&$\ {}   3.180\ {}    $ 
&$\ {}   3.6845\ {}    $ 
&${}\bar\omega_2\ \ \  $
&$\ {}   3.998\ $
&$\ {}   4.32\  {} $ 
&$\ {}   4.524\ {}$ 
&$\ {}   5.313\ {}$ 
\\
$\quad{}     \omega_3\ \ \        $ 
&$\ {}   5.18\  $ 
&$\ {}   5.63 \     $
&$\ {}   5.912\     $ 
&$\ {}   7.038\     $ 
&$\quad{} \bar\omega_3\ \ \   $ 
&$\ {}   6.38\ {} $
&$\ {}   6.96\ {}    $ 
&$\ {}   7.33\ {} $
&$\ {}   8.85\ {} $
\end{tabular}
\end{center}
\begin{center}
\begin{minipage}{.95\hsize}
  \vskip.1cm {\small {\bf Table 3:} $\phi$-even and $\phi$-odd
    eigenvalues for the sharp, quartic, optimised and background field
    cutoff, using $\phi_{\rm max}=0.43, 0.45$, $0.46$, and
    $0.46$, respectively (see text).\\ }
\end{minipage}
\end{center}

Now we proceed with the background field flow, where plain momenta
$q^2$ in the regulator are replaced by $\Gamma^{(2)}[\bar\phi]$, the
full inverse propagator evaluated at some background field $\bar \phi$
\cite{Litim:2002hj} (see also \cite{Litim:2002xm}). Identifying the
background field with the physical mean leads to a partial
diagonalisation, which should further stabilise the flow.  Background
field flows are closely related to the proper-time renormalisation
group of Liao \cite{Liao:1994fp}, to which they reduce once an
additional flow term proportional to $\partial_t \Gamma^{(2)}$ is
dropped.  Implicit to this approach is that differences between
fluctuation and background field are neglected --- an approximation,
which in the present theory becomes exact in the infra-red limit.
Hence, as detailed in \cite{Litim:2002hj}, we expect this
approximation to be viable in the vicinity of a critical point. \step

Here, we use the flow $\partial_t\Gamma_k=\Tr\,
\exp-\Gamma_k^{(2)}/k^2$ to leading order in a derivative
expansion. It is a background field flow in the proper-time
approximation with regulator $R_{\rm bg}$ given by (13), (14) of
\cite{Litim:2002hj}.  Amongst the proper-time flows, it has best
stability properties \cite{Litim:2001hk} (see also
\cite{Litim:2001dt}). This is reflected by the very fast convergence
of $\bar\omega_1$ with the truncation (Fig.~5). We also stress that
the first two eigenvalues in the $\phi$-even sector, which agree with
earlier results in \cite{Mazza:2001bp,Litim:2001hk}, are very close to
the physical values. The further subleading corrections-to-scaling
exponents are increasingly larger than the values for $R_{\rm
opt}$. This trend is indicative for the potential effect of higher
order corrections.\step

Finally we comment on the Polchinski renormalisation group
\cite{Polchinski}. It is related to the flow \eq{ERG} by a Legendre
transform and additional field rescalings. In consequence, both
methods have inequivalent derivative expansions. To leading order, the
Polchinski flow is independent of the regularisation
\cite{Ball:1994ji}, in contradistinction to the present approach,
$e.g.$~\cite{Litim:2002cf}. For $R_{\rm opt}$, critical exponents in
the $\phi$-even sector agree to high precision with those from the
Polchinski flow. The numerical value for the $\phi$-odd eigenvalue
$\bar\omega_1$ given in Tab.~2 for $R_{\rm opt}$, also agrees with
preliminary results from the Polchinski flow \cite{CB}.  If
these findings persist, they confirm the deeper link between the two
methods even for the $\phi$-odd sector.\\[.5ex]

\noindent
{\bf 4. Discussion and conclusion}\step

We have studied symmetric and anti\-symmetric corrections to scaling
at criticality for systems belonging to the $3d$ Ising universality
class. The first six subleading universal corrections-to-scaling
exponents are obtained from an exact renormalisation group. Best
results are achieved for optimised flows, which have enhanced
convergence and stability properties. In addition, we have assessed
the cutoff dependence for smooth, sharp and a background field
cutoff. This study also served as an indicator for higher order
effects. Results from the standard and the background field flows have
to be seen on slightly different footings due to qualitative
differences in the approximations.\step

For the optimised flow, the leading symmetric and antisymmetric
correction-to-scaling exponents are $\omega= 0.6557$ and $\omega_A=
1.886$. For different regularisations ranging from sharp to optimised
cutoffs and including (excluding) the background field flow, the
exponents vary between $\omega\approx 0.60- 0.76$ $(0.60- 0.66)$ and
$\omega_A\approx 1.7 - 2.2$ ($1.7-1.9)$. Higher order corrections due
to a non-vanishing anomalous dimension lead to $\omega\approx 0.8$,
and similar corrections are expected for $\omega_A$.  Expressed in
terms of the exponent $\Delta_A=\omega_A\,\nu$, our results are
$\Delta_A\approx 1.22$ for the optimised, $\Delta_A\approx 1.2$ for
the quartic, $\Delta_A\approx 1.17$ for the sharp and $\Delta_A\approx
1.35$ for the background field cutoff. This compares well with
$\Delta_A\approx 1.3$ which is often assumed in the analysis of
experimental data, $e.g.$~\cite{KWS}. The leading symmetric
corrections to scaling are $\Delta=\omega \,\nu\approx 0.42-0.48$,
increasing towards $\Delta\approx 0.52$ once anomalous dimensions are
taken into account.  This cutoff dependence indicates the expected
size of higher order effects. Our results for $\omega_A$ are
consistent with the estimate $\omega_A> 1.5$ based on Pad\'e
resummation of the $\epsilon$-expansion \cite{wegner}, and with
$\omega_A\approx 2.4$ from the scaling field method
\cite{ScalingField}.  We notice that all sharp cutoff eigenvalues are
systematically smaller than those from any other cutoff. This
reflects, we believe, the notoriously poor convergence behaviour of
sharp cutoff flows.\step

In conclusion, we have established that the leading anti\-symmetric
corrections to scaling are consistently suppressed compared to the
leading symmetric ones. Within the errors, the exponent $\omega_A$ is
more than twice as big as $\omega$. Hence, the scaling
behaviour $\sim L^{-0.5}$ as seen in a Monte-Carlo simulation of the
electro-weak theory clearly dominates over both the leading symmetric
$\sim L^{-\omega}$ and antisymmetric $\sim L^{-\omega_A}$ corrections
to scaling and therefore cannot be explained with the exponent
$\omega_A$.

{\it Acknowledgements:} { LV would like to thank C.~Bervillier for e-mail
correspondence. DFL~thanks the University of Santi\-ago de Chile for
hospitality. This work was supported in part by Fondecyt-CHILE
No.1020061 and No.7020061.}

\end{document}